\documentstyle[12pt]{article}

\begin{document}
\title{Theory of electroweak interactions without spontaneous
symmetry breaking}
\author{Bing An Li\\
Department of Physics and Astronomy, University of Kentucky\\
Lexington, KY 40506, USA}

\maketitle
\begin{abstract}
A electroweak theory without spontaneous symmetry breaking is
studied in this paper.
A new symmetry breaking of $SU(2)_{L}\times U(1)$, axial-vector
symmetry breaking,
caused by the combination of
the axial-vector component of the intermediate boson and the fermion mass
is found
in  electroweak theory.
The mass of the W boson is resulted in the combination of the axial-vector
symmetry breaking and the explicit symmetry breaking by the fermion masses.
The Z boson gains mass from the axial-vector symmetry breaking only.
\(m^{2}_{W}={1\over2}g^{2}m^{2}_{t}\),
\(m^{2}_{Z}=\rho m^{2}_{W}/cos^{2}\theta_{W}\) with
\(\rho\simeq 1\), and
\(G_{F}={1\over 2\sqrt{2}m^{2}_{t}}\)
are obtained.
They are in excellent agreement with data.
The $SU(2)_{L}\times U(1)$ invariant generating functional of the Green's
functions is constructed and the theory is proved to be renormalizable.
The cancellation of the gauge dependent spurious poles and the ghosts is
resulted in the gauge independence of the renormalized s-matrix,
which is proved by using the Ward-Takahashi identity.
\end{abstract}

\newpage
The Standard Model[1] of electroweak interactions is
successful in many aspects. In this model spontaneous symmetry breaking
is introduced to
generate the masses for the W and Z.
So far
Higgs has not been discovered yet. On the other hand, a
Higgs/Hierarchy problem[2] has been revealed.
There are many different attempts[3,4]
trying to solve this problem: W and Z bosons are composite; Higgs
fields are bound states of fermions; supersymmetry etc.
The top quark has been discovered in Fermi laboratory[5], whose mass
has been determined to be
\begin{equation}
m_{t}=173.8\pm 5.2 GeV [6].
\end{equation}
The value of $m_{t}$ is at the same order of magnitude
as the masses of the W and the
Z bosons. As a matter of fact, before the discovery of the top quark
there were attempts of finding the relationship between top quark and
intermediate bosons by using various mechanism[4].
In this paper we propose a new approach in which the masses of W and Z bosons
are generated without spontaneous symmetry breaking.
This paper is organized as: 1) Lagrangian and formalism; 2) model of
axial-vector symmetry breaking; 3) masses of W and Z bosons;
4) renormalizability of the theory;
5) Ward-Takahashi identity and unitarity; 6) Feynman rules; 7) Phenomenology;
8) summary.
\section{Lagrangian and formalism}
The Lagrangian of the Standard Model consists of boson fields,
fermions, and Higgs.
The couplings between fermions and bosons have been extensively tested.
Theoretical results are in excellent agreement with data.
The mass
terms of fermions(except for neutrinos) are well established.
The Lagrangian of the boson fields is constructed by gauge principle.
Therefore,
in the Lagrangian of the Standard Model
the part of the boson fields, the interactions
between fermions and bosons, and the mass terms of fermions are
reliable.
On the other hand,
the Higgs sector of the Lagrangian of the Standard Model
has not been determined yet. In this paper we study the dynamical
properties of the Lagrangian without the Higgs sector[7]
\begin{eqnarray}
\lefteqn{{\cal L}=
-{1\over4}A^{i}_{\mu\nu}A^{i\mu\nu}-{1\over4}B_{\mu\nu}B^{\mu\nu}
+\bar{q}\{i\gamma\cdot\partial-M\}q}
\nonumber \\
&&+\bar{q}_{L}\{{g\over2}\tau_{i}
\gamma\cdot A^{i}+g'{Y\over2}\gamma\cdot B\}
q_{L}+\bar{q}_{R}g'{Y\over2}\gamma\cdot Bq_{R}\nonumber \\
&&+\bar{l}\{i\gamma\cdot\partial-M_{f}\}l
+\bar{l}_{L}\{{g\over2}
\tau_{i}\gamma\cdot A^{i}-{g'\over2}\gamma\cdot B\}
l_{L}-\bar{l}_{R}g'\gamma\cdot B l_{R}.
\end{eqnarray}
Summation over $q_{L}$, $q_{R}$, $l_{L}$, and $l_{R}$ is
implicated respectively in Eq.(2).

It is necessary to point out that in eq.(2) the boson fields are still
elementary fields and
the couplings
between the bosons and the fermions of the Standard Model
remain unchanged, therefore, the success of the Standard Model is
kept.
In the Standard Model the masses of the W and the Z bosons are via
spontaneous symmetry breaking mechanism generated. In this
paper we study whether $m_{W}$ and $m_{Z}$ can be
generated from this Lagrangian(2) without spontaneous symmetry breaking.

Due to
the fermion mass terms the Lagrangian(2) is no longer gauge invariant.
Without losing generality, we study the properties of the
Lagrangaian of the
generation of t and b quarks. The Lagrangian
of this doublet is
\begin{eqnarray}
\lefteqn{{\cal L}=
-{1\over4}A^{i}_{\mu\nu}A^{i\mu\nu}-{1\over4}B_{\mu\nu}B^{\mu\nu}
+\bar{t}\{i\gamma\cdot\partial-m_{t}\}t
+\bar{b}\{i\gamma\cdot\partial-m_{b}\}b}\nonumber \\
&&+\bar{\psi}_{L}\{{g\over2}\tau_{i}
\gamma\cdot A^{i}+g'{1\over6}\gamma\cdot B\}
\psi_{L}
+{2\over3}g'\bar{t}_{R}\gamma\cdot Bt_{R}
-{1\over3}g'\bar{b}_{R}\gamma
\cdot Bb_{R},
\end{eqnarray}
where
\(\psi_{L}=\left(\begin{array}{c}
                t\\b
                 \end{array} \right)_{L}.\)
The quark part of
the Lagrangian(3) is rewritten as
\begin{equation}
{\cal L}=\bar{\psi}\{i\gamma\cdot\partial+\gamma\cdot v+
\gamma\cdot a\gamma_{5}-m\}\psi,
\end{equation}
where $\psi$ is the doublet of t and b quarks,
\(m=\left(\begin{array}{c}
          m_{t}\hspace{1cm}0\\
          0\hspace{1cm}m_{b}
          \end{array} \right),\)
\(v_{\mu}=\tau_{i}v^{i}_{\mu}+\omega_{\mu}\),
\(v^{1,2}_{\mu}={g\over4}A^{1,2}_{\mu}\),
\(v^{3}_{\mu}={g\over4}A^{3}_{\mu}+{g'\over4}B_{\mu}\),
\(\omega_{\mu}={g'\over6}B_{\mu}\),
\(a_{\mu}=\tau_{i}a^{i}_{\mu}\),
\(a^{1,2}_{\mu}=-{g\over4}
A^{1,2}_{\mu}\),
\(a^{3}_{\mu}=-{g\over4}A^{3}_{\mu}+{g'\over4}B_{\mu}\).

The mass term, $\bar{\psi}m\psi$, is not invariant under the
transformation
\[\psi\rightarrow e^{i(\alpha_{1}\tau_{1}+\alpha_{2}
\tau_{2})}\psi.\]
Therefore, the charged gauge symmetry is explicitly broken by the quark
masses and
the charged bosons, W, are expected to gain masses.
On the other hand, the quark mass term is invariant under two gauge
transformations
\[\psi\rightarrow e^{i\alpha}\psi\]
and
\[\psi\rightarrow e^{i\alpha_{3}\tau_{3}}\psi.\]
The theory should have two massless neutral bosons. One of the two
neutral bosons is photon. It is obvious that a new symmetry breaking
is needed to make the Z boson massive.

Using path integral
to integrate out the quark fields, in Euclidean space
the Lagrangian of boson fields
is obtained
\begin{equation}
{\cal L}=lndetD,
\end{equation}
where
\[{\cal D}=\gamma\cdot\partial-i\gamma\cdot v-i\gamma\cdot a\gamma_{5}+m.\]
The real and imaginary parts of the Lagrangian(5) are
\begin{equation}
{\cal L}_{Re}={1\over2}ln det({\cal D}^{\dag}{\cal D}),\;\;\;
{\cal L}_{Im}={1\over2}ln det(\frac{{\cal D}}{{\cal D}^{\dag}}),
\end{equation}
where
\begin{equation}
{\cal D}^{\dag}=-
\gamma\cdot\partial+i\gamma\cdot v-i\gamma\cdot a\gamma_{5}+m.
\end{equation}

It is necessary to point out that
${\cal D}^{\dag}{\cal D}$ is a definite positive operator.
In terms of Schwinger's proper time method[8] ${\cal L}_{Re}$ is
expressed as
\begin{equation}
{\cal L}_{Re}={1\over2}\int d^{D}xTr\int^{\infty}_{0}{d\tau\over\tau}
e^{-\tau{\cal D}^{\dag}{\cal D}}.
\end{equation}
Inserting a complete set of plane waves and subtracting the divergence
at \(\tau=0\), we obtain
\begin{equation}
{\cal L}_{Re}={1\over2}\int d^{D}x
\frac{d^{D}p}{(2\pi)^{D}}Tr\int^{\infty}_{0}{d\tau\over \tau}
\{e^{-\tau{\cal D'}^{\dag}{\cal D}^{'}}-e^{-\tau\Delta_{0}}\},
\end{equation}
where
\begin{eqnarray}
\lefteqn{{\cal D'}=\gamma\cdot\partial+i\gamma\cdot p
-i\gamma\cdot v-i\gamma
\cdot a\gamma_{5}+m,\;\;\;
{\cal D}'^{\dag}=-\gamma\cdot\partial-i\gamma\cdot p+i\gamma\cdot v
-i\gamma\cdot a\gamma_{5}+m,}\nonumber \\
&&{\cal D}'^{\dag}{\cal D}'=\Delta_{0}-\Delta,\;\;\;
\Delta_{0}=p^{2}+m^{2}_{1},\;\;\;
m^{2}_{1}={1\over2}(m^{2}_{t}+m^{2}_{b}),\;\;\;
m^{2}_{2}={1\over2}(m^{2}_{t}-m^{2}_{b}),\nonumber \\
&&\Delta=\partial^{2}-(\gamma\cdot v-\gamma\cdot a\gamma_{5})
(\gamma\cdot v+\gamma\cdot a\gamma_{5})-i\gamma\cdot\partial
(\gamma\cdot v+\gamma\cdot a\gamma_{5})\nonumber \\
&&-i(\gamma\cdot v-\gamma\cdot a
\gamma_{5})\gamma\cdot\partial+2ip\cdot\partial
+2p\cdot(v+a\gamma_{5})
-i[\gamma\cdot v,m]\nonumber \\
&&+i\{\gamma\cdot a,m\}\gamma_{5}
-m^{2}_{2}\tau_{3}.
\end{eqnarray}
After the integration over $\tau$, ${\cal L}_{Re}$ is expressed as
\begin{equation}
{\cal L}_{Re}={1\over2}\int d^{D}x\frac{d^{D}p}{(2\pi)^{D}}
\sum^{\infty}_{n=1}{1\over n}\frac{1}{(p^{2}+m^{2}_{1})^{n}}
Tr\Delta^{n}.
\end{equation}
Due to \({\cal L}_{IM}(-m,-\gamma_{5})=-{\cal L}_{IM}(m,\gamma_{5})\),
${\cal L}_{Im}$ doesn't contribute to the masses of W and Z bosons
at least at the tree level of the boson fields. Therefore, the study
of ${\cal L}_{Im}$ is beyond the scope of this paper..

${\cal L}_{Re}$ is used to investigate
the symmetry breaking mechanism and the masses of the intermediate bosons.
As mentioned above,
the $SU(2)_{L}\times U(1)$ symmetry is explicitly broken by both
quark and lepton masses.
However, another symmetry breaking mechanism is needed.

It is necessary to emphasize on that the field theory of electroweak
interactions is different from $QED$ and $QCD$. In $QED$ and $QCD$
photon and gluons are pure vector fields respectively.
{\bf Due to parity
nonconservation the intermediate bosons
have both vector and
axial-vector components which are written in the forms of
v and a in
Eq.(4).}
In this paper it is shown that this property of the intermediate
boson fields results in another
$SU(2)_{L}\times U(1)$ symmetry breaking.
From the expression of $\Delta$(10)
it is seen that in company with fermion mass
the vector component v of the intermediate boson field
appears in a commutator $[v,m]$, while the axial-vector
component a appears in an anticommutator
$\{a,m\}$. Due to this property
the axial-vector component of boson field causes a new
symmetry breaking, the axial-vector symmetry breaking.
\section{Model of axial-vector symmetry breaking}
In order to show how the axial-vector field results in a symmetry breaking
a model is studied in this section.
The Lagrangian of a vector field and a
fermion($QED$) is
\begin{equation}
{\cal L}=-{1\over4}F_{\mu\nu}F^{\mu\nu}+\bar{\psi}\{i\gamma\cdot
\partial+e\gamma\cdot v\}\psi-m\bar{\psi}\psi.
\end{equation}
This Lagrangian(12) is invariant under the gauge transformation
\[\psi\rightarrow e^{i\alpha(x)}\psi,\;\;\; v_{\mu}\rightarrow
v_{\mu}+{1\over e}\partial_{\mu}\alpha.\]
Using the Eqs.(10,11) and integrating out the fermion field,
the real part of the Lagrangian without derivatives is obtained
\begin{equation}
{\cal L}_{Re}={1\over2}\int d^{D}x\int\frac{d^{D}p}{(2\pi)^{D}}
\sum^{\infty}_{n=1}{1\over n}\frac{1}{(p^{2}+m^{2})^{n}}
Tr(2p\cdot v-v^{2})^{n}.
\end{equation}
Only \(n=1, 2\) contribute to the mass term
\begin{equation}
{\cal L}_{M}=-{D\over2}\int d^{D}x\int\frac{d^{D}p}{(2\pi)^{D}}
\frac{1}{p^{2}+m^{2}}v^{2}+D\int d^{D}x\int\frac{d^{D}p}{(2\pi)^{D}}
\frac{1}
{(p^{2}+m^{2})^{2}}p\cdot vp\cdot v=0.
\end{equation}
These two terms cancel each other. As expected,
no mass for the vector field is generated.

In analogy with Eq.(12) the Lagrangian of an axial-vector field and a
fermion is constructed as
\begin{equation}
{\cal L}=-{1\over4}F_{\mu\nu}F^{\mu\nu}+\bar{\psi}\{i\gamma\cdot
\partial+e\gamma\cdot a\gamma_{5}\}\psi-m\bar{\psi}\psi.
\end{equation}
This Lagrangian(15) is invariant under the gauge transformation
\[\psi\rightarrow e^{i\alpha(x)}\psi,\;\;\; a_{\mu}\rightarrow
a_{\mu}+{1\over e}\partial_{\mu}\alpha\gamma_{5}.\]
Using the Eqs.(10,11) and integrating out the fermion field, the real part of
the Lagrangian without derivatives
is obtained
\begin{equation}
{\cal L}_{Re}={1\over2}\int d^{D}x\int\frac{d^{D}p}{(2\pi)^{D}}
\sum^{\infty}_{n=1}{1\over n}\frac{1}{(p^{2}+m^{2})^{n}}
Tr(2p\cdot a\gamma_{5}-a^{2}+i\{\gamma\cdot a,m\}\gamma_{5})^{n}.
\end{equation}
Only \(n=1, 2\) contribute to the mass term
\begin{eqnarray}
{\cal L}_{M}=-{D\over2}\int d^{D}x\int\frac{d^{D}p}{(2\pi)^{D}}
\frac{1}{p^{2}+m^{2}}a^{2}+D\int d^{D}x\int\frac{d^{D}p}{(2\pi)^{D}}
\frac{1}
{(p^{2}+m^{2})^{2}}\{p\cdot ap\cdot a+m^{2}a^{2}\}\nonumber \\
=\frac{1}{(4\pi)^{2}}D\Gamma(2-{D\over2})m^{2}a_{\mu}a^{\mu}.
\end{eqnarray}
The axial-vector field gains mass, therefore, the symmetry is broken.
This symmetry breaking is called axial-vector symmetry breaking in this paper.

Comparing with eq.(13), in Eq.(16) there is one term more
which is the anticommutator
$i\{\gamma\cdot a, m\}\gamma_{5}$.
Therefore, the theory of
axial-vector field is very different from the vector field.
The symmetry in the theory
of axial-vector field is broken by the combination of the axial-vector
field and the mass of the fermion. The mass of the axial-vector field
is generated by this axial-vector symmetry breaking.
\section{Masses of W and Z bosons}
In terms of the Lagrangian(11) the masses of the
intermediate bosons are calculated.
The terms related to the masses only
is separated from Eq.(11)
\begin{eqnarray}
\lefteqn{{\cal L}_{M}={1\over2}\int d^{D}x\int\frac{d^{D}p}{(2\pi)^{D}}
\sum^{\infty}_{n=1}{1\over n}\frac{1}{(p^{2}+m^{2}_{1})^{n}}Tr\{
-(\gamma\cdot v-\gamma\cdot a\gamma_{5})}\nonumber \\
&&(\gamma\cdot v+\gamma\cdot
a\gamma_{5})+2p\cdot(v+a\gamma_{5})
+i[m,\gamma\cdot v]+i\{m,
\gamma\cdot a\}\gamma_{5}-m^{2}_{2}\tau_{3}\}^{n}.
\end{eqnarray}
The contributions of the fermion masses to $m_{W}$ and $m_{Z}$
are needed to be calculated to all orders.

Four kinds of terms of the Lagrangian(18)
contribute to the masses of the bosons
\begin{eqnarray}
\lefteqn{{\cal L}^{1}=-{1\over2}\int\frac{d^{D}p}{(2\pi)^{D}}\sum^{\infty}
_{n=1}{1\over n}
\frac{1}{(p^{2}+m^{2}_{1})^{n}}Tr(\gamma\cdot v-\gamma\cdot a\gamma_{5})
(\gamma\cdot v+\gamma\cdot a\gamma_{5})(-m^{2}_{2}\tau_{3})^{n-1},}\nonumber \\
&&{\cal L}^{2}=
2\int\frac{d^{D}p}{(2\pi)^{D}}\sum^{\infty}
_{n=2}{1\over n}\frac{(-m^{2}_{2})^{n-2}}{(p^{2}+m^{2}_{1})^{n}}
\sum^{n-2}_{k=0}(n-1-k)Tr
p\cdot(v+a\gamma_{5})\tau_{3}^{k}
p\cdot(v+a\gamma_{5})\tau_{3}^{n-2-k},\nonumber \\
&&{\cal L}^{1+2}={\cal L}^{1}+{\cal L}^{2}=-\frac{8N_{c}}{(4\pi)^{2}}
\sum^{\infty}_{k=1}\frac{1}{(2k+1)(2k-1)}(\frac{m^{2}_{2}}{m^{2}_{1}})
^{2k}({g\over4})^{2}m^{2}_{1}\sum^{2}_{i=1}A^{i}_{\mu}A^{i\mu},\\
&&{\cal L}^{3}=-{1\over2}\int\frac{d^{D}p}{(2\pi)^{D}}\sum^{\infty}
_{n=2}{1\over n}\frac{(-m^{2}_{2})^{n-2}}{(p^{2}+m^{2}_{1})^{n}}\sum^{n-2}
_{k=0}(n-1-k)Tr[\gamma\cdot v,m]\tau_{3}^{k}[\gamma\cdot
v,m]\tau_{3}^{n-2-k}\nonumber \\
&&=\frac{2N_{c}}{(4\pi)^{2}}D\Gamma(2-{D\over2})({g\over4})^{2}
m^{2}_{-}\sum^{2}_{i=1}A^{i}_{\mu}A^{i\mu}\nonumber \\
&&+\frac{8N_{C}}{(4\pi)^{2}}\sum^{\infty}_{k=2}\frac{1}
{(2k-1)(2k-2)}(\frac{m^{2}_{2}}{m^{2}_{1}})^{2k-2}({g\over4})^{2}
m^{2}_{-}\sum^{2}_{i=1}A^{i}_{\mu}A^{i\mu},\\
&&{\cal L}^{4}={1\over2}\int\frac{d^{D}p}{(2\pi)^{D}}\sum^{\infty}
_{n=2}{1\over n}\frac{(m^{2}_{2})^{n-2}}{(p^{2}+m^{2}_{1})^{n}}\sum^{n-2}
_{k=0}(n-1-k)Tr\{\gamma\cdot a,m\}\tau_{3}^{k}
\{\gamma\cdot
a,m\}\tau_{3}^{n-2-k}\nonumber \\
&&=\frac{2N_{c}}{(4\pi)^{2}}D\Gamma(2-{D\over2})\{
({g\over4})^{2}
m^{2}_{+}\sum^{2}_{i=1}A^{i}_{\mu}A^{i\mu}
+m^{2}_{1}[({g\over4})^{2}+({g'\over4})^{2}]Z_{\mu}Z^{\mu}\}\nonumber \\
&&-\frac{8N_{C}}{(4\pi)^{2}}\sum^{\infty}_{k=1}\frac{1}
{(2k-1)2k}(\frac{m^{2}_{2}}{m^{2}_{1}})^{2k-1}\{({g\over4})^{2}
+({g'\over4})^{2}\}
m^{2}_{2}Z_{\mu}Z^{\mu}\nonumber \\
&&+\frac{8N_{C}}{(4\pi)^{2}}\sum^{\infty}_{k=1}\frac{1}
{(2k+1)2k}(\frac{m^{2}_{2}}{m^{2}_{1}})^{2k}({g\over4})^{2}
m^{2}_{+}\sum^{2}_{i=1}A^{i}_{\mu}A^{i\mu},
\end{eqnarray}
where \(m_{+}={1\over2}(m_{t}+m_{b})\) and \(m_{-}={1\over2}
(m_{t}-m_{b})\).

The Eqs.(19-21) show that all the four terms contribute to
the mass of the W boson.
These results indicate that the mass of the charged boson, W,
originates in both the explicit $SU(2)_{L}\times U(1)$
symmetry breaking by the fermion masses and the axial-vector symmetry
breaking caused by the combination of the axial-vector component
and the fermion mass($\{a_{\mu},m\}$).
Only ${\cal L}^{4}$ contributes to $m_{Z}$. ${\cal L}^{4}$ is related to $
\{a_{\mu},m\}$. Therefore, $m_{Z}$ is generated by the axial-vector
symmetry breaking.
As expected, a U(1) symmetry
remains and the neutral vector meson, the photon,
is massless.

It is found that the series of the fermion masses are convergent.
Putting all the four terms(19-21) together, the Lagrangian of
the masses of W and Z
are obtained in Minkowski space
\begin{eqnarray}
\lefteqn{{\cal L}_{M}={1\over2}\frac{N_{C}}{(4\pi)^{2}}\{{D\over4}
\Gamma(2-{D\over2})(4\pi{\mu^{2}\over m^{2}_{1}})
^{{\epsilon\over2}}+{1\over2}[1-ln(1-x)-
(1+{1\over x}){\sqrt{x}\over2}
ln\frac{1+\sqrt{x}}{1-\sqrt{x}}]\}m^{2}_{1}g^{2}
\sum^{2}_{i=1}A^{i}_{\mu}A^{i\mu}}\nonumber \\
&&+{1\over2}\frac{N_{C}}{(4\pi)^{2}}\{{D\over4}
\Gamma(2-{D\over2})(4\pi{\mu^{2}\over m^{2}_{1}})
^{{\epsilon\over2}}-{1\over2}[ln(1-x)+\sqrt{x}
ln\frac{1+\sqrt{x}}{1-\sqrt{x}}]\}m^{2}_{1}(g^{2}+g'^{2})
Z_{\mu}Z^{\mu},
\end{eqnarray}
where $N_{C}$ is the number of colors and
\(x=({m^{2}_{2}\over m^{2}_{1}})^{2}\). It is necessary to point
out that
\begin{equation}
a^{3}_{\mu}={1\over4}\sqrt{g^{2}+g'^{2}}Z_{\mu}.
\end{equation}
In the same way, other two generations of quarks,
\(\left(\begin{array}{c}
         u\\d
        \end{array} \right) \)
and
\(\left(\begin{array}{c}
         c\\s
        \end{array} \right) \),
and three generations of leptons
\(\left(\begin{array}{c}
         \nu_{e}\\e
        \end{array} \right) \),
\(\left(\begin{array}{c}
         \nu_{\mu}\\\mu
        \end{array} \right) \), and
\(\left(\begin{array}{c}
         \nu_{\tau}\\\tau
        \end{array} \right) \)
contribute to the masses of W and Z bosons too. By changing the
definitions of $m^{2}_{1}$ and x to the quantities of other
generations in Eq.(22),
the contributions of the other two quark generations
are found. Taking off the factor $N_{C}$ and changing $m^{2}_{1}$ and
x to corresponding quantities of leptons, the contributions of the
leptons to $m_{W}$ and $m_{Z}$ are obtained.
In this paper the effects of CKM matrix are not taken into account.
The final expressions of the masses of W and Z bosons
are the sum of the contributions of the three quark and the three
lepton generations.
It is learned from the processes deriving Eq.(22) that
\begin{enumerate}
\item Due to the U(1) symmetry the neutral vector
field $sin\theta_{W}A^{3}_{\mu}+cos\theta_{W}B_{\mu}$(photon
field) is massless;
\item The W boson gains mass from both the explicit and the axial-vector
$SU(2)_{L}\times U(1)$ symmetry breaking;
\item
$m_{Z}$ is resulted in the axial-vector symmetry breaking only.
\end{enumerate}

In order to renormalize the boson fields it is needed to study the kinetic
terms of the boson fields.
Up to all orders of fermion masses,
the kinetic terms of the intermediate boson fields
are obtained from the Lagrangian(11). For the generation of t and b quarks
there are nine terms
\begin{eqnarray}
\lefteqn{{\cal L}^{1}=-{1\over2}\int\frac{d^{D}p}{(2\pi)^{D}}\sum^{\infty}
_{n=3}{1\over n}\frac{(-m^{2}_{2})^{n-3}}{(p^{2}+m^{2}_{1})^{n}}\sum
^{n-3}_{k=0}\sum^{n-3-k}_{k_{1}=0}(n-2-k-k_{1})Tr[\gamma\cdot v,m]
\partial^{2}\tau^{k+k_{1}}_{3}[\gamma\cdot v,m]\tau^{n-3-k-k_{1}}
}\nonumber \\
&&=\frac{8N_{C}}{(4\pi)^{2}}\frac{m^{2}_{-}}{m^{2}_{1}}\sum^{\infty}
_{k=1}\frac{1}{(2k+1)(2k-1)}(\frac{m^{2}_{2}}{m^{2}_{1}})^{2k-2}
({g\over4})^{2}\sum^{2}_{i=1}A^{i}_{\mu}\partial^{2}A^{i\mu},\\
&&{\cal L}^{2}=2\int\frac{d^{D}p}{(2\pi)^{D}}\sum^{\infty}
_{n=4}{1\over n}\frac{(-m^{2}_{2})^{n-4}}{(p^{2}+m^{2}_{1})^{n}}\sum
^{n-4}_{k=0}\sum^{n-4-k}_{k_{1}=0}\sum^{n-4-k-k_{1}}_{k_{2}=0}
(n-3-k-k_{1}-k_{2})\nonumber \\
&&Tr[\gamma\cdot v,m]\tau^{k+k_{1}+k_{2}}_{3}(p\cdot \partial)^{2}
[\gamma\cdot v,m]\tau^{n-4-k-k_{1}-k_{2}}
\nonumber \\
&&=-\frac{4N_{C}}{(4\pi)^{2}}\frac{m^{2}_{-}}{m^{2}_{1}}\sum^{\infty}
_{k=1}\frac{1}{(2k+1)(2k-1)}(\frac{m^{2}_{2}}{m^{2}_{1}})^{2k-2}
({g\over4})^{2}\sum^{2}_{i=1}A^{i}_{\mu}\partial^{2}A^{i\mu},\\
&&{\cal L}^{3}={1\over2}\int\frac{d^{D}p}{(2\pi)^{D}}\sum^{\infty}
_{n=3}{1\over n}\frac{(-m^{2}_{2})^{n-3}}{(p^{2}+m^{2}_{1})^{n}}\sum
^{n-3}_{k=0}\sum^{n-3-k}_{k_{1}=0}(n-2-k-k_{1})\nonumber \\
&&Tr\{\gamma\cdot a,m\}
\partial^{2}\tau^{k+k_{1}}_{3}\{\gamma\cdot a,m\}\tau^{n-3-k-k_{1}}
\nonumber \\
&&=\frac{8N_{C}}{(4\pi)^{2}}\frac{m^{2}_{+}}{m^{2}_{2}}\sum^{\infty}
_{k=1}\frac{1}{(2k+1)(2k-1)}(\frac{m^{2}_{2}}{m^{2}_{1}})^{2k-1}
({g\over4})^{2}\sum^{2}_{i=1}A^{i}_{\mu}\partial^{2}A^{i\mu}+
\frac{8N_{C}}{(4\pi)^{2}}{1\over3}a^{3}_{\mu}\partial^{2}a^{3\mu},\\
&&{\cal L}^{4}=-2\int\frac{d^{D}p}{(2\pi)^{D}}\sum^{\infty}
_{n=4}{1\over n}\frac{(-m^{2}_{2})^{n-4}}{(p^{2}+m^{2}_{1})^{n}}\sum
^{n-4}_{k=0}\sum^{n-4-k}_{k_{1}=0}\sum^{n-4-k-k_{1}}_{k_{2}=0}
(n-3-k-k_{1}-k_{2})\nonumber \\
&&Tr\{\gamma\cdot a,m\}\tau^{k+k_{1}+k_{2}}_{3}(p\cdot \partial)^{2}
\{\gamma\cdot a,m\}\tau^{n-4-k-k_{1}-k_{2}}
\nonumber \\
&&=-\frac{4N_{C}}{(4\pi)^{2}}\frac{m^{2}_{+}}{m^{2}_{2}}\sum^{\infty}
_{k=2}\frac{1}{(2k-1)(2k-3)}(\frac{m^{2}_{2}}{m^{2}_{1}})^{2k-3}
({g\over4})^{2}\sum^{2}_{i=1}A^{i}_{\mu}\partial^{2}A^{i\mu}
-{1\over3}\frac{4N_{C}}{(4\pi)^{2}}a^{3}_{\mu}\partial^{2} a^{3\mu},\\
&&{\cal L}^{1+2+3+4}=\frac{4N_{C}}{(4\pi)^{2}}\sum^{\infty}_{k=1}
\frac{1}{(2k+1)(2k-1)}(\frac{m^{2}_{1}}{m^{2}_{2}})^{2k-2}
({g\over4})^{2}\sum^{2}_{i=1}A^{i}_{\mu}\partial^{2}A^{i\mu}+
\frac{4N_{C}}{3(4\pi)^{2}}a^{3}_{\mu}\partial^{2}a^{3\mu},\nonumber \\
&&{\cal L}^{5}=-{1\over2}\int\frac{d^{D}p}{(2\pi)^{D}}\sum^{\infty}
_{n=2}{1\over n}\frac{(-m^{2}_{2})^{n-2}}{(p^{2}+m^{2}_{1})^{n}}
\sum^{n-2}_{k=0}(n-1-k)Tr(\gamma\cdot v-\gamma\cdot a\gamma_{5})
\partial^{2}\tau^{k}_{3}(\gamma\cdot v+\gamma\cdot
a\gamma_{5})\tau^{n-2-k}	_{3}\nonumber \\
&&=-\frac{N_{C}}{(4\pi)^{2}}{D\over4}\Gamma(2-{D\over4})Tr(v_{\mu}\partial	^{2}v
^{\mu}+a_{\mu}\partial^{2}a^{\mu})\nonumber \\		
&&-\frac{4N_{C}}{(4\pi)^{2}}\sum^{\infty}_{k=2}\frac{1}{(2k-1)(2k-2)}
(\frac{m^{2}_{2}}{m^{2}_{1}})^{2k-2}({g\over4})^{2}A^{i}_{\mu}\partial^{2}
A^{i\mu}\nonumber \\
&&-\frac{2N_{C}}{(4\pi)^{2}}\sum^{\infty}_{k=2}\frac{1}{2k-2}(\frac{m^{2}_{2}}{
m^{2}_{1}})^{2k-2}\{v^{3}_{\mu}\partial^{2}v^{3\mu}+\omega_{\mu}
\partial^{2}\omega^{\mu}+a^{3}_{\mu}\partial^{2}a^{3\mu}\}\nonumber \\
&&+\frac{4N_{C}}{(4\pi)^{2}}\sum^{\infty}_{k=1}\frac{1}{2k-1}(\frac{m^{2}_{2}}{m
^{2}_{1}})^{2k-1}
\omega_{\mu}\partial^{2}v^{3\mu},\\
&&{\cal L}^{6}=2\int\frac{d^{D}p}{(2\pi)^{D}}\sum^{\infty}
_{n=3}{1\over n}\frac{(-m^{2}_{2})^{n-3}}{(p^{2}+m^{2}_{1})^{n}}\sum
^{n-3}_{k=0}\sum^{n-2-k}_{k_{1}=0}(n-2-k-k_{1})\nonumber \\
&&Trp\cdot(v+a\gamma_{5})\tau^{k
+k_{1}}_{3}\partial^{2}p\cdot(v+a\gamma_{5})\tau^{n-3-k-k_{1}}\nonumber \\
&&={2\over3}\frac{N_{C}}{(4\pi)^{2}}{D\over4}\Gamma(2-{D\over4})Tr(v_{\mu}
\partial^{2}v^{\mu}
+a_{\mu}\partial^{2}a^{\mu})\nonumber \\		
&&+\frac{8N_{C}}{(4\pi)^{2}}\sum^{\infty}_{k=2}\frac{1}{(2k+1)(2k-1)(2k-2)}
(\frac{m^{2}_{2}}{m^{2}_{1}})^{2k-2}({g\over4})^{2}A^{i}_{\mu}\partial^{2}
A^{i\mu}\nonumber \\
&&+\frac{4N_{C}}{3(4\pi)^{2}}\sum^{\infty}_{k=2}\frac{1}{2k-2}(\frac{m^{2}_{2}}{
m^{2}_{1}})^{2k-2}\{v^{3}_{\mu}\partial^{2}v^{3\mu}+\omega_{\mu}
\partial^{2}\omega^{\mu}+a^{3}_{\mu}\partial^{2}a^{3\mu}\}\nonumber \\
&&-\frac{8N_{C}}{3(4\pi)^{2}}\sum^{\infty}_{k=1}\frac{1}{2k-3}(\frac{m^{2}_{2}}
{m^{2}_{1}})^{2k-3}
\omega_{\mu}\partial^{2}v^{3\mu},\\
&&{\cal L}^{7}=-8\int\frac{d^{D}p}{(2\pi)^{D}}\sum^{\infty}
_{n=4}{1\over n}\frac{(-m^{2}_{2})^{n-4}}{(p^{2}+m^{2}_{1})^{n}}\sum
_{k,k_{1},k_{2}}(n-3-k-k_{1}-k_{2})\nonumber \\
&&Trp\cdot(v+a\gamma_{5})\tau^{k+k_{1}+k_{2}}
(p\cdot\partial)^{2}p\cdot(v+a\gamma_{5})\tau^{n-4-k-k_{1}-k_{2}}
\nonumber \\
&&=-{1\over3}\frac{N_{C}}{(4\pi)^{2}}{D\over4}\Gamma(2-{D\over4})Tr(v_{\mu}
\partial^{2}v^{\mu}-2\partial_{\mu}v^{\mu}\partial_{\nu}v^{\nu}
+a_{\mu}\partial^{2}a^{\mu}-2\partial_{\mu}a^{\mu}\partial_{\nu}a^{\nu})
\nonumber \\	
&&-\frac{4N_{C}}{(4\pi)^{2}}\sum^{\infty}_{k=3}\frac{1}{(2k-1)(2k-3)(2k-4)}
(\frac{m^{2}_{2}}{m^{2}_{1}})^{2k-2}({g\over4})^{2}(A^{i}_{\mu}\partial^{2}
A^{i\mu}-2\partial_{\mu}A^{i\mu}\partial_{\nu}A^{i\nu})\nonumber \\
&&-\frac{2N_{C}}{3(4\pi)^{2}}\sum^{\infty}_{k=3}\frac{1}{2k-4}(\frac{m^{2}_{2}}
{m^{2}_{1}})^{2k-4}\{v^{3}_{\mu}\partial^{2}v^{3\mu}-2\partial
_{\mu}v^{3\mu}\partial_{\nu}v^{3\nu}
+\omega_{\mu}\partial^{2}\omega^{\mu}\nonumber \\
&&-2\partial_{\mu}\omega^{\mu}\partial_{\nu}\omega^{\nu}
+a^{3}_{\mu}\partial^{2}a^{3\mu}-2\partial_{\mu}a^{3\mu}\partial_{\nu}
a^{3\nu}\}\nonumber \\
&&+\frac{4N_{C}}{3(4\pi)^{2}}\sum^{\infty}_{k=2}\frac{1}{2k-3}(\frac{m^{2}_{2}}
{m^{2}_{1}})^{2k-3}(
\omega_{\mu}\partial^{2}v^{3\mu}-2\partial_{\mu}\omega^{\mu}
\partial_{\nu}v^{3\nu}),\\
&&{\cal L}^{8}=2\int\frac{d^{D}p}{(2\pi)^{D}}\sum^{\infty}
_{n=3}{1\over n}\frac{(-m^{2}_{2})^{n-3}}{(p^{2}+m^{2}_{1})^{n}}\sum
^{n-3}_{k=0}\sum^{n-3-k}_{k_{1}=0}(n-2-k-k_{1})\nonumber \\
&&Trp\cdot(v+a\gamma_{5})\tau
^{k+k_{1}}_{3}p\cdot\partial\gamma\cdot\partial(\gamma\cdot v+\gamma\cdot a
\gamma_{5})\tau^{n-3-k-k_{1}},\nonumber  \\	
&&={2\over3}\frac{N_{C}}{(4\pi)^{2}}{D\over4}\Gamma(2-{D\over4})Tr(v^{\mu}
\partial_{\mu\nu}v^{\nu}+a^{\mu}\partial_{\mu\nu}a^{\nu})\nonumber \\
&&+\frac{8N_{C}}{(4\pi)^{2}}\sum^{\infty}_{k=2}\frac{1}{(2k+1)(2k-1)(2k-2)}
(\frac{m^{2}_{2}}{m^{2}_{1}})^{2k-2}({g\over4})^{2}(A^{i\mu}\partial
_{\mu\nu}A^{i\nu})\nonumber \\
&&+\frac{4N_{C}}{3(4\pi)^{2}}\sum^{\infty}_{k=2}\frac{1}{2k-2}(\frac{m^{2}_{2}}
{m^{2}_{1}})^{2k-2}\{v^{3\mu}\partial_{\mu\nu}
v^{3\nu}
+\omega^{\mu}\partial_{\mu\nu}\omega^{\nu}+a^{3\mu}\partial_{\mu\nu}
a^{3\nu})\nonumber \\
&&-\frac{8N_{C}}{3(4\pi)^{2}}\sum^{\infty}_{k=2}\frac{1}{2k-3}(\frac{m^{2}_{2}}
{m^{2}_{1}})^{2k-3}
\omega^{\mu}\partial_{\mu\nu}v^{3\nu},\\
&&{\cal L}^{9}=2\int\frac{d^{D}p}{(2\pi)^{D}}\sum^{\infty}
_{n=3}{1\over n}\frac{(-m^{2}_{2})^{n-3}}{(p^{2}+m^{2}_{1})^{n}}\sum
^{n-3}_{k=0}\sum^{n-3-k}_{k_{1}=0}(n-2-k-k_{1})\nonumber \\
&&Tr
(\gamma\cdot v-\gamma\cdot a\gamma_{5})\gamma\cdot\partial p\cdot\partial
\tau
^{k+k_{1}}_{3}p\cdot(\gamma\cdot v+\gamma\cdot a
\gamma_{5})\tau^{n-3-k-k_{1}},\nonumber  \\	
&&={\cal L}^{8}.
\end{eqnarray}

Taking other two generations of quarks and three generations of
leptons into account in Eqs.(24-32) and adding them  together,
in Minkowski space the kinetic terms are obtained
\begin{eqnarray}
\lefteqn{{\cal L}_{K}=-{1\over4}\sum_{i=1,2}
(\partial_{\mu}A^{i}_{\nu}-
\partial_{\nu}A^{i}_{\mu})^{2}\{1+{1\over(4\pi)^{2}}g^{2}
\sum_{q,l}N[{D\over12}
\Gamma(2-{D\over2})(4\pi)^{{\epsilon\over2}}({\mu^{2}\over m^{2}_{1}})
^{{\epsilon\over2}}-{1\over6}+f_{1}]\}}\nonumber \\
&&-{1\over4}(\partial_{\mu}A^{3}_{\nu}-\partial_{\nu}A^{3}_{\mu})^{2}
\{1+\frac{1}{(4\pi)^{2}}g^{2}\sum_{q,l}N
[{D\over12}\Gamma(2-{D\over2})
(4\pi)^{{\epsilon\over2}}({\mu^{2}\over m^{2}_{1}})
^{{\epsilon\over2}}-{1\over6}+{1\over6}f_{2}]\}\nonumber \\
&&-{1\over4}(\partial_{\mu}B_{\nu}-\partial_{\nu}B_{\mu})^{2}
\{1+\frac{N_{C}}{(4\pi)^{2}}g'^{2}\sum_{q}[{11\over9}
{D\over12}\Gamma(2-{D\over2})
(4\pi)^{{\epsilon\over2}}({\mu^{2}\over m^{2}_{1}})
^{{\epsilon\over2}}-{1\over6}+{11\over54}f_{2}-{1\over18}f_{3}
]\}\nonumber \\
&&+\frac{1}{(4\pi)^{2}}g'^{2}\sum_{l}[
{D\over4}\Gamma(2-{D\over2})
(4\pi)^{{\epsilon\over2}}({\mu^{2}\over m^{2}_{1}})
^{{\epsilon\over2}}-{1\over6}+{1\over2}f_{2}+{1\over6}f_{3}
]\}\nonumber \\
&&-\frac{1}{(4\pi)^{2}}{gg'\over12}(\partial_{\mu}A^{3}_{\nu}
-\partial_{\nu}A^{3}_{\mu})(\partial^{\mu}B^{\nu}-\partial^{\nu}B^{\mu})
\{N_{G}-{2\over3}N_{C}\sum_{q}f_{3}+2\sum_{l}f_{3}\},
\end{eqnarray}
where $\sum_{q}$ and $\sum_{l}$ stand for summations of
generations of quarks and leptons respectively,
\(N=N_{C}\) for q and \(N=1\) for lepton, \(N_{G}=3N_{C}+3\),
x depends on fermion generation and is defined in Eq.(22),
\begin{eqnarray}
\lefteqn{f_{1}={4\over9}-{1\over6x}
-{1\over6}ln(1-x)
+{1\over4\sqrt{x}}({1\over3x}-1)
ln\frac{1+\sqrt{x}}{1-\sqrt{x}},}\nonumber \\
&&f_{2}=-ln(1-x),\;\;\;
f_{3}=
{1\over2}{1\over\sqrt{x}}ln\frac{1+\sqrt{x}}
{1-\sqrt{x}}.
\end{eqnarray}
Following results are obtained from Eq.(33)
\begin{enumerate}
\item The boson fields and the coupling constants g and g'
have to be redefined by multiplicative renormalization
\begin{eqnarray}
A^{i}_{\mu}\rightarrow (Z^{A}_{1})^{{1\over2}}A^{i}_{\mu},
B_{\mu}\rightarrow (Z^{B}_{2})^{{1\over2}}B{\mu},\nonumber \\
g_{1}=(Z^{A}_{1})^{-{1\over2}}g,
g_{2}=(Z^{B}_{2})^{-{1\over2}}g',
\end{eqnarray}
where
\begin{eqnarray}
\lefteqn{Z^{A}_{1}=
1+{1\over(4\pi)^{2}}g^{2}\sum_{q,l}N
[{D\over12}
\Gamma(2-{D\over2})(4\pi)^{{\epsilon\over2}}
({\mu^{2}\over m^{2}_{1}})
^{{\epsilon\over2}}-{1\over6}+f_{1}]}\nonumber \\
&&Z^{B}_{2}=1+\frac{N_{C}}{(4\pi)^{2}}g'^{2}\sum_{q}[{11\over9}
{D\over12}\Gamma(2-{D\over2})
(4\pi)^{{\epsilon\over2}}({\mu^{2}\over m^{2}_{1}})
^{{\epsilon\over2}}-{1\over6}+{11\over54}f_{2}-{1\over18}f_{3}
]\nonumber \\
&&+\frac{1}{(4\pi)^{2}}g'^{2}\sum_{l}[
{D\over4}\Gamma(2-{D\over2})
(4\pi)^{{\epsilon\over2}}({\mu^{2}\over m^{2}_{1}})
^{{\epsilon\over2}}-{1\over6}+{1\over2}f_{2}+{1\over6}f_{3}
].
\end{eqnarray}
\item There is a crossing term between
$A^{3}_{\mu}$ and $B_{\mu}$, which is written as
\begin{eqnarray}
\lefteqn{g_{1}g_{2}(\partial_{\mu}A^{3}_{\nu}-\partial_{\nu}A^{3}_{\mu})
(\partial_{\mu}B_{\nu}-\partial_{\nu}B_{\mu})
=e^{2}(\partial_{\mu}A_{\nu}-\partial_{\nu}A_{\mu})^{2}
-e^{2}(\partial_{\mu}Z_{\nu}-\partial_{\nu}Z_{\mu})^{2}}
\nonumber \\
&&+e(g_{1}cos\theta_{W}-g_{2}sin\theta_{W})
(\partial_{\mu}A_{\nu}-\partial_{\nu}A_{\mu})
(\partial_{\mu}Z_{\nu}-\partial_{\nu}Z_{\mu}),
\end{eqnarray}
where \(sin\theta_{W}=\frac{g_{2}}{\sqrt{g^{2}_{1}+g^{2}_{2}}}\),
\(cos\theta_{W}=\frac{g_{1}}{\sqrt{g^{2}_{1}+g^{2}_{2}}}\),
and \(e=\frac{g_{1}g_{2}}{\sqrt{g^{2}_{1}+g^{2}_{2}}}\).
Therefore, the photon and the Z fields are needed to be renormalized
again
\begin{equation}
(1+{\alpha\over4\pi}f_{4})^{{1\over2}}A_{\mu}\rightarrow A_{\mu},\;\;\;
(1-{\alpha\over4\pi}f_{4})^{{1\over2}}Z_{\mu}\rightarrow Z_{\mu},
\end{equation}
where $\alpha$ is the fine structure constant and
\[f_{4}={1\over3}N_{G}-{2\over3}\sum_{q}f_{3}+{2\over3}\sum_{l}f_{3}.\]
After these renormalizations(35,39),
${\cal L}_{K}$(33) is rewritten as
\begin{eqnarray}
\lefteqn{{\cal L}_{K}=
-{1\over 4}(\partial_{\mu}A_{\nu}-\partial_{\nu}
A_{\mu})^{2}
-{1\over 4}\sum_{i=1,2}(\partial_{\mu}A^{i}_{\nu}-\partial_{\nu}
A^{i}_{\mu})^{2}
-{1\over 4}(\partial_{\mu}Z_{\nu}-\partial_{\nu}
Z_{\mu})^{2}}\nonumber \\
&&-{1\over4}\frac{\alpha}{4\pi}
({g_{1}\over g_{2}}-{g_{2}\over g_{1}})
(1-\frac{\alpha^{2}}{(4\pi)
^{2}}f^{2}_{4})^{-{1\over2}}f_{4}
(\partial_{\mu}A_{\nu}-
\partial_{\nu}A_{\mu})(\partial^{\mu}Z^{\nu}-\partial^{\nu}Z^{\mu}).
\end{eqnarray}
\item
The interaction between photon and Z boson is predicted in Eq.(39).
\end{enumerate}

Now we can study the values of the masses of W and Z bosons.
After the renormalizations(35,38) there are still divergences in the
mass formulas of $m_{W}$ and $m_{Z}$(22).
In Eq.(22) the fermion masses are bare
physical quantities. It is reasonable to redefine the fermion
masses by multiplicative renormalization
\begin{eqnarray}
Z_{m}m^{2}_{1}=m^{2}_{1,P},\nonumber \\
Z_{m}=\frac{N}{(4\pi)^{2}}\{N_{G}{D\over4}\Gamma(2-{D\over2})
(4\pi)^{{\epsilon\over2}}({\mu^{2}\over m^{2}_{1}})
^{{\epsilon\over2}}+{1\over2}[1-ln(1-x)-
(1+{1\over x}){\sqrt{x}\over2}
ln\frac{1+\sqrt{x}}{1-\sqrt{x}}]\},
\end{eqnarray}
for each generation of fermions. The index "P" is omitted in the rest
of the paper.
Now the mass of W boson is obtained from Eq.(22)
\begin{equation}
m^{2}_{W}={1\over2}g^{2}\{m^{2}_{t}+m^{2}_{b}+m^{2}_{c}+m^{2}_{s}
+m^{2}_{u}+m^{2}_{d}+m^{2}_{\nu_{e}}+m^{2}_{e}
+m^{2}_{\nu_{\mu}}+m^{2}_{\mu}+m^{2}_{\nu_{\tau}}+m^{2}_{\tau}\}
\end{equation}
Obviously, the top quark mass dominates the $m_{W}$
\begin{equation}
m_{W}={g\over\sqrt{2}}m_{t}.
\end{equation}
Using the values \(g=0.642\) and \(m_{t}=173.8\pm5.2 GeV\)[6], it is found
\begin{equation}
m_{W}=78.9\pm2.4 GeV,
\end{equation}
which is in excellent agreement with data $80.41\pm0.10$GeV[6].
The Fermi coupling constant is derived from eq.(42)
\[G_{F}=\frac{1}{2\sqrt{2}m^{2}_{t}}=1.03\times10^{-5}m^{-2}_{N},\]
where \(m_{t}=173.8GeV\) is taken.

Using Eqs.(22,35,38),
the mass formula of the Z boson
is written as
\begin{equation}
m^{2}_{Z}=\rho m^{2}_{W}(1+{g^{2}_{2}\over g^{2}_{1}}),
\end{equation}
where
\begin{eqnarray}
\lefteqn{\rho=(1-{\alpha\over4\pi}f_{4})^{-1}\sum_{q,l}N
\{{D\over4}
\Gamma(2-{D\over2})(4\pi)
^{{\epsilon\over2}}({\mu^{2}\over m^{2}_{1}})
^{{\epsilon\over2}}-{1\over2}[ln(1-x)+\sqrt{x}
ln\frac{1+\sqrt{x}}{1-\sqrt{x}}]\}}
\nonumber \\
&&/\sum_{q,l}N\{{D\over4}
\Gamma(2-{D\over2})(4\pi{\mu^{2}\over m^{2}_{1}})
^{{\epsilon\over2}}+{1\over2}[1-ln(1-x)-
(1+{1\over x}){\sqrt{x}\over2}
ln\frac{1+\sqrt{x}}{1-\sqrt{x}}]\}
\end{eqnarray}
Comparing
with the infinites in Eqs.(45),
we have
\begin{equation}
m^{2}_{Z}=\rho m^{2}_{W}/cos^{2}\theta_{W},\;
\rho=(1-{\alpha\over4\pi}f_{4})^{-1},
\end{equation}
Due to the smallness of the factor ${\alpha\over4\pi}$
in the reasonable ranges of the quark masses
and the upper
limits of neutrino masses we expect
\begin{equation}
\rho\simeq 1.
\end{equation}
Therefore,
\begin{equation}
m_{Z}=m_{W}/cos\theta_{W}
\end{equation}
is a good approximation. Eq.(48) is the prediction of the minimum
Standard Model.

Introduction of a cut-off leads to
\begin{equation}
{D\over 4}
\Gamma(2-{D\over2})(4\pi)^{{\epsilon\over2}}({\mu^{2}\over m^{2}_{1}})
^{{\epsilon\over2}}\rightarrow
ln(1+{\Lambda^{2}\over
m^{2}_{1}})-1+\frac{1}{1+{\Lambda^{2}\over m^{2}_{1}}}.
\end{equation}
Taking $\Lambda\rightarrow\infty$,
Eq.(46) is obtained.
On the other hand,
the cut-off might be estimated by the value of the $\rho$
factor. The cut-off can be
considered as
the energy scale of unified electroweak theory.
\section{Renormalizability of the theory}
It has been proved that gauge invariant gauge field theory is renormalizable[9].
However, the Lagrangian(2) is not gauge invariant. We need to study whether
the theory described by Eq.(2) is renormalizable.
It is well known that some filed theories with symmetry breaking  are
renormalizable too. The $\sigma$ model[10] is heuristic to the theory studied
in this paper. The Lagrangian of the $\sigma$ model is
written as[10]
\[{\cal L}={\cal L}_{sym}+c\sigma,\]
where
\[{\cal L}_{sym}={1\over2}[(\partial_{\mu}\pi)^{2}+(\partial_{\mu}\sigma)^{2}]
-{1\over2}\mu^{2}(\pi^{2}+\sigma^{2})-{1\over4}(\pi^{2}+\sigma^{2})^{2}\]
is invariant under $U(1)$ transformation. However, the $U(1)$ symmetry is
broken by a term of dimension one, $c\sigma$. It has been shown[10] that
the Green's functions of this model are generated by the generating
functional of the symmetric theory. The latter is
given by
\[W[J]=exp\{iZ[J]\}=\int[d\sigma][d\pi]exp\{i\int d^{4}x[{\cal L}_{sym}(x)
+J_{\sigma}(x)\sigma(x)+J_{\pi}(x)\pi(x)]\},\]
where $J_{\sigma}\sigma+J_{\pi}\pi$ is U(1) invariant.
The Green's functions are calculated in the limits
\[J_{\sigma}=c,\;\;\;J_{\pi}=0.\]
The $\sigma$ model is renormalizable[10].

It has been proved[11] that
spontaneous broken gauge theories are renormalizable too.

The Lagrangian(2) studied in this paper is not gauge invariant and it can
be rewritten as
\begin{equation}
{\cal L}={\cal L}_{sym}-\bar{q}Mq-\bar{l}M_{f}l,
\end{equation}
where ${\cal L}_{sym}$ is the gauge invariant part of the Lagrangian(2).
The $SU(2)_{L}\times U(1)$ symmetry of this theory is broken by operators
of dimension three. The generating functional of the Green's functions is
constructed by introducing the corresponding source terms.
Following Ref.[11], the generating functional is
constructed as
\begin{eqnarray}
\lefteqn{W_{F}[J]=exp\{iZ[J]\}=\int[dA^{i}_{\mu}][dB_{\mu}][d\psi_{L}]
[d\bar{\psi_{L}}][d\psi_{R}][d\bar{\psi_{R}}]\Delta_{F_{1}}[A^{i}_{\mu}
]\Delta_{F_{2}}[B_{\mu}]}\nonumber \\
&&exp[i\int d^{4}x\{{\cal L}_{s}-{1\over2}F^{T}_{1}F_{1}
-{1\over2}F^{2}_{2}+J^{i}_{\mu}A^{i\mu}
+J_{\mu}B_{\mu}\nonumber \\
&&+\bar{J}_{\psi_{L}}\psi_{L}+\bar{\psi_{L}}J_{\psi_{L}}
+\bar{J}_{\psi_{R}}\psi_{R}
+\bar{\psi_{R}}J_{\psi_{R}}\}],
\end{eqnarray}
where
\[F_{1j}(A^{i}_{\mu})-a_{j}=0\]
is the general gauge condition of the $A^{i}_{\mu}$ fields,
j is the index of the group $SU_{L}(2)
$, $a_{j}$ is a constant, $F_{2}$ is the gauge condition of the $B_{\mu}$
field, and
\[{\cal L}_{s}={\cal L}_{sym}+{\cal L}_{m}.\]
It is similar to the $\sigma$ model that
${\cal L}_{m}$ is introduced as the external source terms of
the fermion masses. As an
example, for the t and b quark generation it is
\begin{equation}
{\cal L}^{tb}_{m}=\bar{\psi}^{tb}_{L}t_{R}J^{t}_{c}+J^{t\dag}_{c}\bar{t}
\psi^{tb}_{L}+\bar{\psi}^{tb}_{L}b_{R}J^{b}+J^{b\dag}\bar{b}_{R}
\psi^{tb}_{L},
\end{equation}
where $J^{t,b}$ are external sources and they are doublets of $SU(2)_{L}$
and $J^{t}_{c}$ is the dual
representation of $J^{t}$. For $J^{t}$ and $J^{b}$,
the weak hypercharge \(Y=-1,1\)
respectively. Therefore, ${\cal L}^{tb}_{m}$ is invariant under
$SU(2)_{L}\times U(1)$.
In the same way other generations can be added.
All Green's functions can be derived from $W_{F}[J]$ under the gauge conditions
$F_{1,2}$ in the limit
\begin{eqnarray}
\lefteqn{J^{i}_{\mu}=0,\;\;\;J_{\mu}=0,\;\;\;J_{\psi_{L,R}}=0,
\;\;\;J_{\bar{\psi}_{L,R}}=0,}\nonumber \\
&&J^{t}_{c}=\left(\begin{array}{c}
                   -m_{t}\\0
                 \end{array}  \right) \nonumber \\
&&J^{b}=\left(\begin{array}{c}
                   0\\-m_{b}
                 \end{array}  \right)
\end{eqnarray}
Using these limits, the mass terms of fermions are obtained
\begin{equation}
-m_{t}\bar{t}t-m_{b}\bar{b}b-....
\end{equation}
For the sake of simplicity, in the study above the CKM matrix has not been
taken into account. There is no
problem to put the CKM matrix in.

${\cal L}_{s}$ is gauge invariant.
The expression of the generating functional $W_{F}[J]$(51)
of the Green's functions of
this theory is in the same form as the one of renormalizable gauge field
theories[9].
Therefore, the theory studied in this paper is
renormalizable.

It is necessary to point out that if the mass terms of the W and the Z bosons
are added to the Lagrangain(2) there is no way to make the theory
renormalizable.

The renormalizations of the gauge fields and fermion fields
are
\begin{eqnarray}
\lefteqn{A^{i}_{0\mu}=(Z^{A}_{1})^{{1\over2}}A^{i}_{\mu},}\nonumber \\
&&B_{0\mu}=(Z^{B}_{2})^{{1\over2}}B_{\mu},\nonumber \\
&&\psi^{L}_{0j}=(Z^{j}_{L})^{{1\over2}}\psi^{L}_{j},\nonumber \\
&&\psi^{R}_{0j\sigma}=(Z_{R}^{j\sigma})^{{1\over2}}\psi^{R}_{j\sigma},\nonumber
\\
&&g=Z^{A}_{1}(Z^{A}_{2})^{-{3\over2}}g_{1},\nonumber \\
&&g^{'}=Z^{B}_{2}(Z^{B}_{2})^{-{3\over2}}g
_{2},
\end{eqnarray}
and the renormalizations for $J^{t}_{c}$, $J^{b}$, and other external sources.
\section{Ward-Takahashi identity and unitarity}
In Ref.[11] the Ward-Takahashi identity has been used to prove that the
renormalized s-matrix is independent of the gauge. Then
it follows that the poles in the propagators are spurious
and should be cancelled each other. The theory is proved to be unitary.
Following Ref.[11], the same proof can be done.

$W_{F}[J]$ is invariant under $SU(2)_{L}\times U(1)$. Therefore, there are
two Ward-Takahashi identities.
Following Ref.[11], the Ward-Takahashi identities of the theory proposed in
this paper can be derived. The Ward-Takahashi identity corresponding to
$SU(2)_{L}$ invariance is derived as
\begin{equation}
\{-F_{1i}({1\over i}{\delta\over\delta J})+J_{a}(\Gamma^{j}_{ab}{1\over i}
{\delta\over\delta J_{b}}+\Lambda^{j}_{a})[M^{-1}_{F_{1}}({1\over i}
{\delta\over\delta J})]_{ji}\}W_{F_{1}}[j]=0,
\end{equation}
where \(a=(\mu l)\) and \(b=(\nu k)\), $\mu$ and $\nu$ are Lorentz indices,
i,j,k,l are the indices of the $SU(2)_{L}$,
\begin{equation}
\Gamma^{j}_{(\mu l)(\nu k)}=-i\epsilon_{lkj}g_{\mu\nu},\;\;\;\Lambda^{j}_{(\mu
i)}=-\delta_{ij}\partial_{\mu},
\end{equation}
which are obtained from the gauge transformation of $SU(2)_{L}$
\begin{equation}
(A^{g}_{\mu})^{i}=A^{i}_{\mu}-i\epsilon_{ijk}u_{j}A^{k}_{\mu}-\partial_{\mu}
u_{i},
\end{equation}
$F_{i}$ is the gauge condition and
\begin{equation}
M_{ij}=\frac{\partial F_{i}}{\partial u_{j}},\;\;\; det M=\Delta_{F}.
\end{equation}
The Ward-Takahashi identity for $U(1)$ is well known and can be derived in
the same way. Eq.(56) is the same as the one presented in Ref.[11].

Following Ref.[11], the Ward-Takahashi identities are used to prove that
\begin{equation}
s=s_{F}/\Pi_{l}(Z_{F})^{{1\over2}}_{l}
\end{equation}
is independent of the gauge condition F, where $s_{F}$ is the s-matrix
calculated
in the gauge condition of F(F means both $F_{1,2}$), $(Z_{F})_{l}$ is the wave
function
renormalization constant for the lth external line, and $\Pi$ extends all over
the external lines.

It is well known that the ghost fields are associated with the matrix $M_{F}$
which is determined from the gauge condition (59). The spurious poles in
the
propagators of W and Z bosons come from the gauge term
$-{1\over2}F^{T}_{1}F_{1}$ in
the generating functional $W_{F}[J]$(51).
Therefore, the spurious poles and the ghosts are gauge dependent.
On the other hand,
it has been shown[11]  that the effects of
the gauge conditions on s-matrix is a redefinition of the renormalization of
the unrenormalized s-matrix[60]. After renormalization the s-matrix
is independent of the gauge condition. Therefore, in principle the spurious
poles and ghosts should be cancelled out in the s-matrix after renormalization.
As claimed in Ref.[11], this cancellation leads to that the theory studied
in this paper is unitary.

\section{Feynman rules}
It has been shown in this paper that the masses of W and Z bosons are obtained
from the fermion loop diagrams. The Lagrangian of the theory
proposed in this paper is written as
\begin{equation}
{\cal L}={\cal L}_{0}+{\cal L}_{i}+{\cal L}_{c},
\end{equation}
where ${\cal L}_{0}$ is the free part of the Lagrangian
\begin{eqnarray}
\lefteqn{{\cal L}_{0}=-{1\over2}(\partial_{\mu}W^{+}_{\nu}-\partial_{\nu}
W^{+}_{\mu})(\partial_{\mu}W^{-}_{\nu}-\partial_{\nu}W^{-}_{\mu})
-{1\over4}(\partial_{\mu}Z_{\nu}-\partial_{\nu}Z_{\mu})^{2}
-{1\over4}F^{\gamma}_{\mu\nu}F^{\gamma\mu\nu}}\nonumber \\
&&+m_{W}^{2}W^{+}_{\mu}W^{-\mu}
+{1\over2}m^{2}_{Z}Z_{\mu}Z^{\mu}\nonumber \\
&&+\bar{\psi_{q}}\{i\partial\cdot\gamma-M\}\psi_{q}
+\bar{\psi_{l}}\{i\partial\cdot\gamma-M_{l}\}\psi_{l},
\end{eqnarray}
${\cal L}_{i}$ is the interaction part and ${\cal L}_{c}$
is the counter part of the Lagrangian. In the ${\cal L}_{c}$ there are two
additional terms
\begin{equation}
-m^{2}_{W}W^{+}_{\mu}W^{-\mu}-{1\over2}m^{2}_{Z}Z_{\mu}Z^{\mu}
\end{equation}
and
\begin{equation}
-\bar{\psi_{q}}\{i\partial\cdot\gamma-M\}\psi_{q}
-\bar{\psi_{l}}\{i\partial\cdot\gamma-M_{l}\}\psi_{l}.
\end{equation}
The Eq.(63) is used to cancel the mass terms of W and Z generated from the
fermion loop calculation and the Eq.(64) is to cancel the fermion mass terms
obtained from the corresponding external source terms in $W_{F}[J]$(51,53,54).

Using the $R_{\xi}$ gauge, the propagators of W and Z bosons are derived as
\begin{equation}
D_{\mu\nu}(k^{2})={i\over k^{2}-m^{2}+i\epsilon}\{-g_{\mu\nu}+(1-
{1\over\xi})\frac{k_{\mu}k_{\nu}}{k^{2}-{m^{2}\over\xi}
+i\epsilon}\}.
\end{equation}
The vertices can all be derived. As a matter of fact, all the propagators,
the vertices, and the renormalization constants up to one loop can be found
in Ref.[12] by getting rid of Higgs and the parameters related to
spontaneous symmetry breaking.
\section{Phenomenology}
This theory keeps the success of the Standard Model. On the other hand, the
whole Higgs sector of the SM does not exist in this theory.
This difference leads to new phenomenology of electroweak interactions.
Detailed
phenomenological study is beyond
the scope of this paper.
\section{Summary}
In the Lagrangian(2) the boson fields are elementary fields. The Lagrangian
of boson fields, the couplings between the fermions and the bosons, and
the fermion mass terms are the same as in the Standard Model.
Therefore, the success of the
Standard Model is kept.
The W and Z
fields are different from photon and gluons, they have both vector and
axial-vector components. It is shown in this paper that
axial-vector field is very different from vector field. A new
symmetry breaking, axial-vector symmetry breaking,
caused by the combination of the axial-vector component
and fermion mass is found. In this theory there are both the explicit symmetry
breaking by fermion masses and the axial-vector symmetry breaking. The charged
boson, W, gains mass from both the symmetry breaking, while the mass of
the neutral boson, Z, originates in the axial-vector symmetry breaking only.
Due to the U(1) symmetry and the vector nature the photon remains
massless. Upon the scheme of multiplicative renormalization the values of
$m_{W}$ and $m_{Z}$ are determined. They are in excellent agreement with
data. The masses of W and Z are obtained without the spontaneous symmetry
breaking.

On the other hand,
following Refs.[9,10,11], the theory is proved to be renormalizable.
The Ward-Takahashi identity is derived and used to prove that the
renormalized s-matrix
is gauge independent and the spurious poles and ghosts should be cancelled out.
The theory is unitary. The Feynman rules of this theory are presented.

The author wishes to thank W.A.Bardeen, M.Chanowitz, and J.Rosner
for discussion.
This research was partially
supported by DOE Grant No. DE-91ER75661.

\end{document}